\documentclass[english, aps, prd, superscriptaddress, preprintnumbers,
nofootinbib, twocolumn]{revtex4-1}
\usepackage[T1]{fontenc}
\usepackage[utf8]{inputenc}
\usepackage{verbatim}
\usepackage{empheq}
\usepackage{amssymb}
\usepackage{amsmath}
\usepackage{amsfonts}
\usepackage{graphicx}
\usepackage{epstopdf}

\usepackage[colorlinks,linktocpage=true]{hyperref}
\hypersetup{citecolor=blue,linkcolor=blue,urlcolor=blue}

\usepackage{babel}
\usepackage{setspace}

\newcommand{\emax}{E_\textrm{max}}
\newcommand{\eqspace}{\hphantom{{}={}}}
\newcommand{\Tr}[1]{\textrm{Tr}{\left(#1\right)}}
\newcommand{\FLM}[2]{F_\textrm{LM}^{\textrm{nf}}{\left(#1 \, | \, #2\right)}}
\newcommand{\FLV}[2]{F_\textrm{LV}^{\textrm{nf}}{\left(#1 \, | \, #2\right)}}
\newcommand{\FLVfin}[2]{F_\textrm{LV,fin}^{\textrm{nf}}{\left(#1 \, |
		\, #2\right)}}
\newcommand{\FLVfinlo}[1]{F_\textrm{LV,fin}^{\textrm{nf}}{\left(#1\right)}}
\newcommand{\FLVlo}[1]{F_\textrm{LV}^{\textrm{nf}}{\left(#1\right)}}
\newcommand{\FLMlo}[1]{F_\textrm{LM}^{\textrm{nf}}{\left(#1\right)}}
\newcommand{\dq}[1]{[\textrm{d}p_{#1}]\, }
\newcommand{\asmu}{{\frac{\tilde \alpha_s}{2\pi}}}
\newcommand{\asmusq}{{\bigg(\hspace{-1.5pt}\frac{\tilde \alpha_s}{2\pi}
		\hspace{-1.5pt}\bigg)^{\hspace{-2pt}2}}}
\newcommand{\FLVV}[1]{F_\textrm{LVV}^{\textrm{nf}}{\left(#1\right)}}
\renewcommand{\Re}{\textrm{Re}}

\begin{document}

\preprint{TTP23-014,~P3H-23-027}

\def\BNL{High Energy Theory Group, Physics Department,
	Brookhaven National Laboratory, Upton, NY 11973, USA}
\def\KIT{Institute for Theoretical Particle Physics,
	Karlsruhe Institute of Technology, 76128 Karlsruhe, Germany}

\title{On the non-factorizable corrections to Higgs boson production in weak
	boson fusion}

\author{Konstantin~Asteriadis}
\email[Electronic address: ]{kasteriad@bnl.gov}
\affiliation{\BNL}

\author{Christian~Br\o{}nnum-Hansen}
\email[Electronic address: ]{christian.broennum-hansen@partner.kit.edu}
\affiliation{\KIT}

\author{Kirill~Melnikov}
\email[Electronic address: ]{kirill.melnikov@kit.edu}
\affiliation{\KIT}

\begin{abstract}
	\noindent
	We discuss the non-factorizable corrections to Higgs boson production in weak
	boson fusion at the Large Hadron Collider. Such corrections depend
	on the finite part of the two-loop virtual amplitude $q \, Q \rightarrow q^\prime
	\, Q^\prime + H$ which, up to now, has only been computed in the eikonal
	approximation.  We combine this contribution with real-virtual and double-real
	non-factorizable QCD corrections and study  their impact on the various
	observables in weak boson fusion. We find that the non-factorizable corrections
	are strongly dominated by the two-loop virtual contributions, while all other
	contributions play a very minor role. This striking imbalance between real and
	virtual contributions is caused by a process-specific kinematic suppression of
	the former and a particular enhancement of the virtual corrections related to a
	Glauber phase.
\end{abstract}

\maketitle

\section{Introduction}
\label{sec:introduction}

Weak boson fusion (WBF) is an important Higgs boson production channel; it has
the second-largest cross section at the Large Hadron Collider (LHC). In addition, it is directly
sensitive to the couplings of the Higgs boson to $W$ and $Z$ bosons allowing for
a detailed exploration of their strengths and Lorentz structures.

Theoretical predictions for Higgs boson production in weak boson fusion are
very advanced. They include  next-to-leading order (NLO)
QCD~\cite{Figy:2003nv,*Berger:2004pca,*Figy:2004pt} and
electroweak~\cite{Ciccolini:2007jr,*Ciccolini:2007ec,*Figy:2010ct} corrections
as well as next-to-next-to-leading order (NNLO)
QCD~\cite{Bolzoni:2010xr,*Bolzoni:2011cu,*Cacciari:2015jma,
	*Cruz-Martinez:2018rod,Asteriadis:2021gpd,Asteriadis:2022ebf} and
next-to-next-to-next-to-leading order (N${}^3$LO) QCD \cite{Dreyer:2016oyx}
corrections.   In addition, effects of multijet merging and an interplay between
fixed order perturbative computations and parton showers in weak boson
fusion was studied in Ref.~\cite{Chen:2021phj}.
However, available QCD corrections are computed in the so-called
factorization approximation where strong interactions between the incoming quark
lines are systematically ignored.

Historically, non-factorizable corrections were neglected because they are
colour-suppressed~\cite{Bolzoni:2010xr,*Bolzoni:2011cu} and, moreover, they
appear at NNLO QCD for the first time. However, it was pointed out in
Ref.~\cite{Liu:2019tuy} that these corrections receive a peculiar
$\pi^2$-enhancement associated with a Glauber phase. In
Refs.~\cite{Liu:2019tuy,Dreyer:2020urf} the numerical impact of
non-factorizable corrections on various observables in WBF was investigated. It
was found that these corrections are somewhat smaller than the factorizable
corrections at NNLO QCD but that they certainly exceed the magnitude of
N${}^3$LO QCD corrections.

To make further progress in understanding the non-factorizable effects in weak
boson fusion, there are two directions to take. First, one can extend the
calculation of the non-factorizable two-loop amplitude for the WBF process $q \ Q
\rightarrow q^\prime \ Q^\prime + H$ beyond the eikonal approximation. This is a
formidable task since it requires the computation of two-loop five-point
amplitudes with two massive propagators and an additional external massive
particle which is  beyond the current state of the art.  Second, one can study
the effects of all the other contributions relevant for computing the
non-factorizable correction through NNLO in perturbative QCD while accounting
for the double-virtual contribution in the eikonal approximation. This is what
we do in this paper.

Computation of NNLO QCD corrections to WBF requires double-real and real-virtual
contributions, in addition to the two-loop virtual corrections. Individually,
each of these contributions is infrared divergent; to properly define them a
subtraction procedure is needed. Since in the past decade remarkable
progress in the development of NNLO QCD subtraction schemes for collider
processes has been made, and since certain features of the non-factorizable
correction to Higgs boson fusion in WBF make the infrared structure of this
process simple, construction of the subtraction scheme for computing the
non-factorizable corrections to WBF becomes straightforward. In fact, the
relevant computation can be borrowed, almost verbatim, from a similar
computation of the non-factorizable corrections to single-top production
reported recently in Ref.~\cite{Bronnum-Hansen:2022tmr}.

It is worth pointing out that the situation with real-virtual contributions is
somewhat peculiar. Although the relevant one-loop amplitudes can be extracted
from an existing computation of NLO QCD corrections to $H+j$ production in weak
boson fusion~\cite{Campanario:2013fsa}, the fact that the corresponding
\emph{six-point} amplitude needs to be evaluated close to singular limits makes
its use in the computation of NNLO QCD corrections non-trivial.

The remaining part of the paper is organized as follows. In the next section we
recapitulate the construction of the infrared-finite fully-differential cross
section suitable for numerical computation. We discuss the numerical implementation and address difficulties with
evaluating subtracted real-virtual contributions in
Section~\ref{sec:implementation}. We then present the results of our
computation and show that the non-factorizable corrections are strongly
dominated by two-loop virtual corrections. We conclude in
Section~\ref{sec:conclusions}.

\section{Construction of an infrared finite cross section}
\label{sec:computation}

\begin{figure}
	\centering
	\hspace{-15pt}\includegraphics[height=75pt]{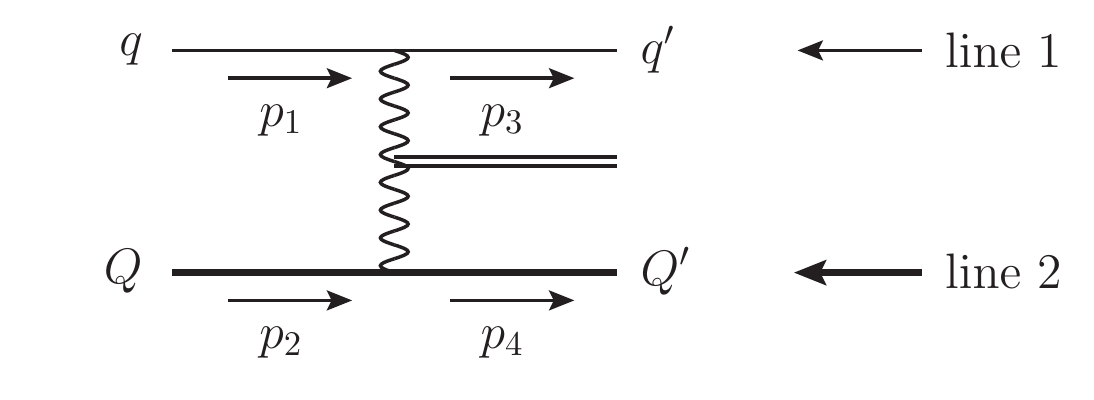}
	\caption{Momentum, parton and line conventions at Born level used throughout
		the discussion. We do not show fermion flow because $q$ and $Q$ each
		represent any (light) quark or anti-quark.}
	\label{fig:conventions}
\end{figure}

A NNLO QCD computation requires the
construction of an infrared-finite cross section which can be integrated over
phase space of final-state particles in four dimensions. This requires the use
of a subtraction scheme since contributions with different number of final-state
partons are not separately finite.

The construction of such a subtraction scheme for the case of non-factorizable
contributions to single-top production was recently presented in
Ref.~\cite{Bronnum-Hansen:2022tmr}. The discussion in that reference applies
almost verbatim to the computation of non-factorizable corrections to Higgs
boson production in weak boson fusion. Because of that, we confine ourselves to
reviewing the major building blocks of such a construction in this section,
and note that further details can be found in
Ref.~\cite{Bronnum-Hansen:2022tmr}.

Non-factorizable corrections involve exchanges of real and virtual gluons
between the two quark lines of the partonic process $q \, Q \rightarrow q^\prime \,
Q^\prime + H$ , where $q$ and $Q$ are arbitrary quarks or anti-quarks, see
Fig.~\ref{fig:conventions}. Such corrections do not contribute at
next-to-leading order due to colour conservation. Indeed,  both real and
virtual non-factorizable corrections at NLO QCD contain just one single colour
generator $T^{a}$ on each fermion line. When one computes the interference of
the one-loop virtual amplitude with the leading-order amplitude or the square of
the real-emission amplitude, the corrections vanish since the colour generators
are traceless.\footnote{ We neglect identical-flavour contributions which are
	known to be suppressed both kinematically and by colour at NLO
	QCD~\cite{Figy:2003nv}.}

\begin{figure*}
	\centering
	\includegraphics[height=79pt]{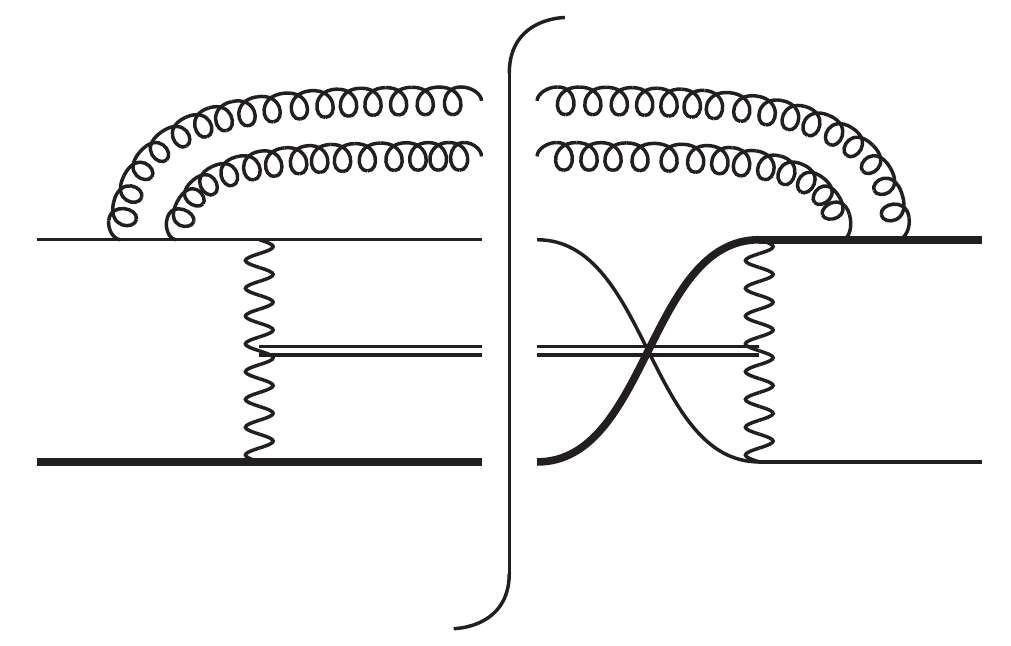}\hspace{0pt}
	\includegraphics[height=79pt]{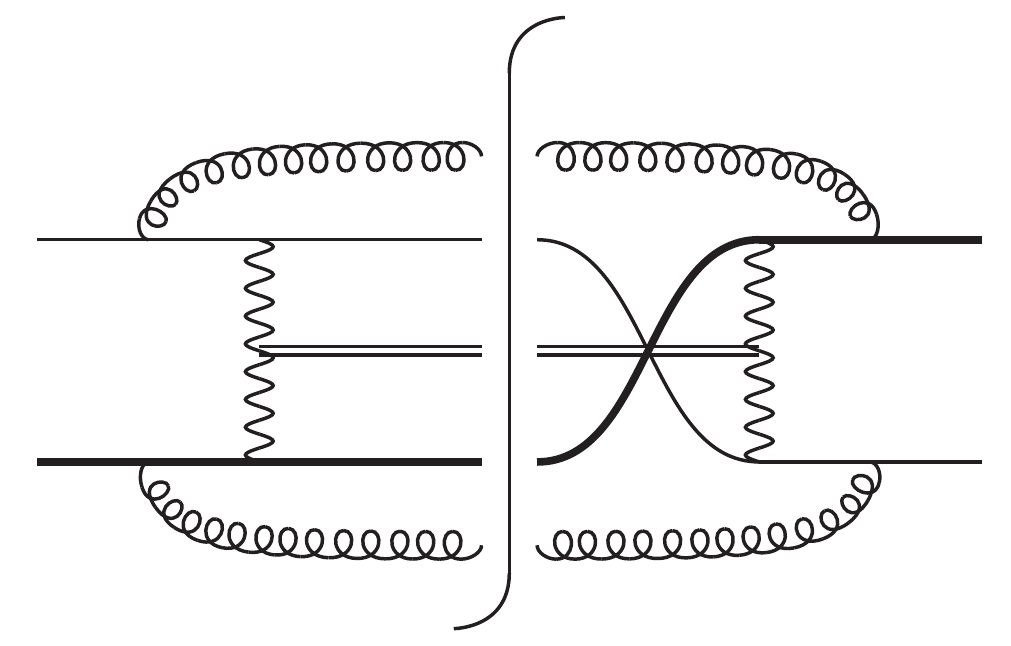}\hspace{0pt}
	\includegraphics[height=79pt]{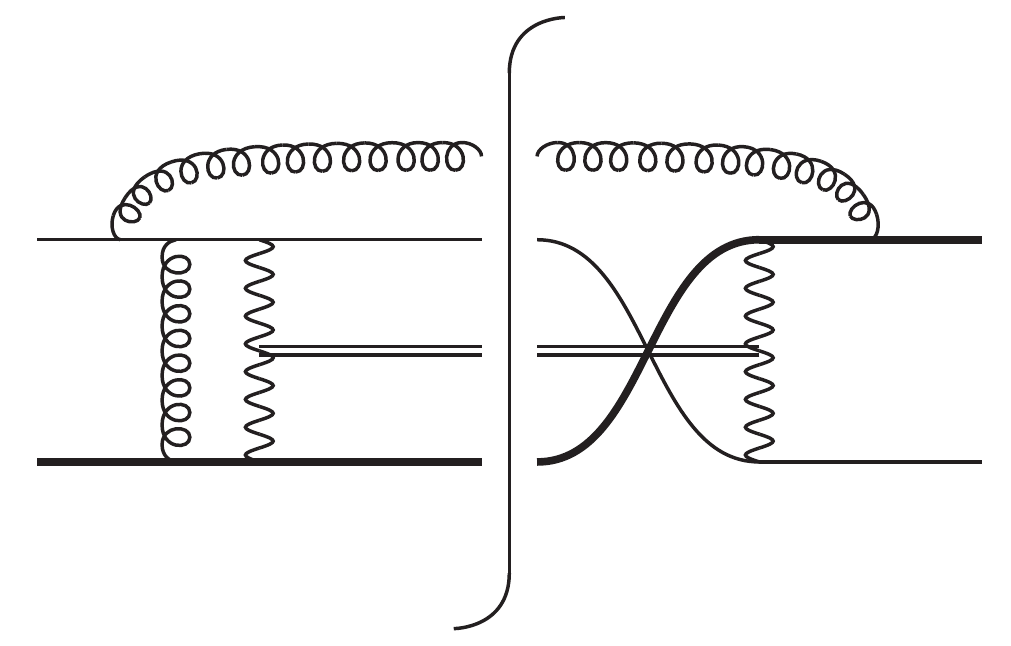} \hspace{0pt}
	\includegraphics[height=79pt]{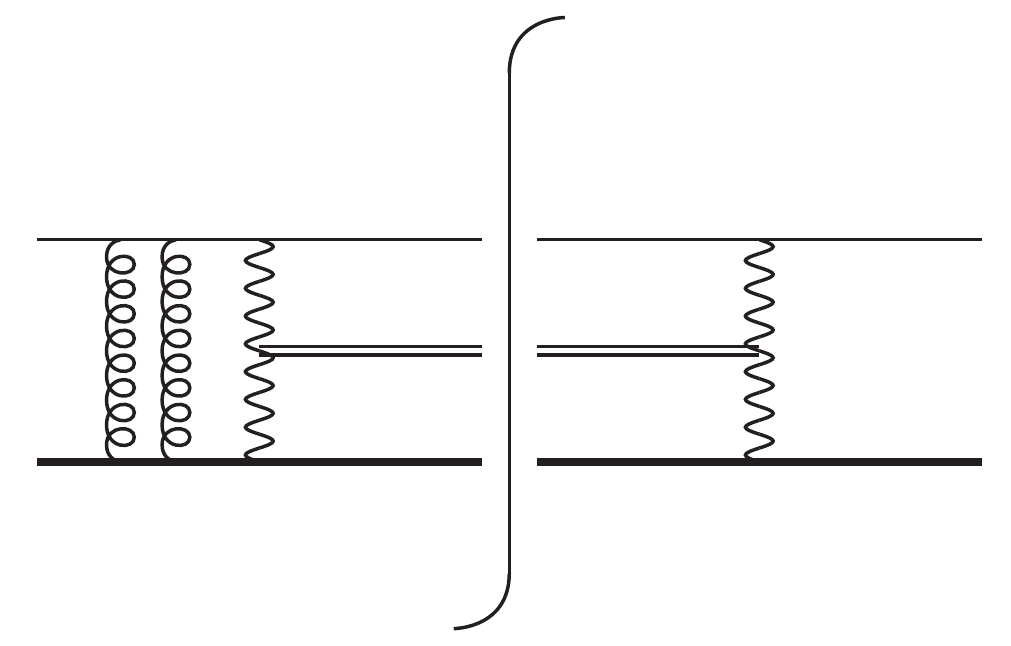} \\[5pt]
	(a) \hspace{110pt} (b) \hspace{110pt} (c) \hspace{110pt} (d)
	\caption{Schematic examples of non-vanishing contributions to the
		non-factorizable double-real (a--b), real-virtual (c), and double-virtual
		(d) amplitude squared. It is easy to see that the colour factor for each
		contribution is $T_R^2 (N_c^2-1)$, as stated in the main text. To
		distinguish  the two \textit{massless} quark lines, one is printed in
		bold.}
	\label{fig:interference}
\end{figure*}

Despite being absent at lower orders, non-factorizable contributions do appear
at NNLO in perturbative QCD. For example, virtual contributions with two gluons
connecting the upper and lower quark lines lead to a colour factor $\Tr{T^a T^b}
= T_R \, \delta^{ab}$ for each line and clearly do not vanish when the
interference with the leading-order amplitude is computed. We show some of the
non-vanishing contributions in Fig.~\ref{fig:interference}. Furthermore, it is
easy to see that non-factorizable contributions at NNLO cannot involve
non-abelian QCD vertices. This feature renders all non-factorizable corrections
QED-like and leads, as we will discuss later in more detail, to a simple
infrared structure of such contributions. We will now consider the various
contributions to the NNLO QCD non-factorizable corrections and review the
construction of the subtraction terms.

\subsection*{Double-real emission contribution}
\label{sec:comp:double-real}

We begin with the \emph{non-factorizable} contributions to the double-real
emission process
\begin{align}
	\begin{split}
		&q(p_1) + Q(p_2) \\
		&\rightarrow q^\prime(p_3) + Q^\prime(p_4) + g(p_5) + g(p_6) + H(p_H) \, .
	\end{split}
\end{align}
All such contributions to the \emph{amplitude squared} carry the same
colour factor given by
\begin{align}
	\label{eq.2}
	\sum_{a,b} \Tr{T^a T^b}^2 = T_R^2 (N_c^2-1) \, ,
\end{align}
where $T_R = 1/2$, $N_c = 3$, $a$ and $b$ are the colour indices of gluons $p_5$
and $p_6$, respectively, and the summation over quark colours has been
performed. Since the colour factor is always the same, it is convenient to work
with \textit{colour-stripped} amplitudes and restore the overall colour factor
at the end.

We write the relevant colour-stripped amplitudes as\footnote{Dependence of the
	amplitude on the Higgs boson momentum $p_H$ is not shown because it is not
	relevant for the present discussion.}
\begin{align}
	A_0^{ij}(1_q,2_Q,3_{q^\prime},4_{Q^\prime}\,|\,5_g,6_g) \, ,
\end{align}
where superscript $i(j) \in \{1,2\}$ refers to one of the two quark lines
from which gluon $5$($6$) is emitted (see Fig.~\ref{fig:conventions}). We
emphasize again that only abelian diagrams contribute to $A_0^{ij}$ and that, to
obtain them, the colour generators in quark-gluon vertices are to be removed.
Similarly, we define \textit{colour-stripped} amplitudes $A_0^i$ for a single
gluon emission from line $i \in \{1,2\}$, and $A_0$ for the amplitude of the
process without additional gluons.

Following Ref.~\cite{Caola:2017dug} we define
\begin{align}
	&\FLM{1_q,2_Q,3_{q^\prime},4_{Q^\prime}}{5_g,6_g} \equiv \mathcal{N} \int
	\textrm{dLips}_{34H} \nonumber \\
	& \times \hat{\mathcal{O}}{(\{p_{i=1,\dots,6},p_H\})} (2\pi)^d \,
	\delta^{(d)}{\bigg(p_1 + p_2 - p_H - \sum_{i=3}^6 p_i\bigg)} \nonumber \\
	\label{eq:rr:flm}
	& \times
	2\,\Re{\left[A_0^{11} {A_{0}^{22}}^\star + A_0^{12} {A_{0}^{21}}^\star
		\right]}(1,2,3,4\, | \, 5,6) \, ,
\end{align}
where dLips${}_{34H}$ is the Lorentz-invariant phase space of the two
final-state fermions and the Higgs boson, $\mathcal{N} = 1/(4N_c^2)$ includes
spin and colour-averaging factors, $\hat{\mathcal{O}}{(\{p_{i=1,\dots,6},
	p_H\})}$ is an arbitrary infrared-safe observable, and $d = 4 -2\epsilon$ is the space-time dimension.

To obtain the partonic differential cross section we restore colour charges
and write
\begin{align}
	\textrm{d}\sigma_\textrm{rr}^\textrm{nf} = \frac{T_R^2(N_c^2-1)}{2s}\;
	\big\langle \FLM{1,2,3,4}{5,6} \big\rangle,
\end{align}
where $s = 2 p_1 \cdot p_2$. We also define $\big\langle \FLM{1,2,3,4}{5,6}
\big\rangle$ as an integral over the two-gluon phase space\footnote{We choose to
	order gluon emissions in energy and, therefore, do not include the factor
	$1/2!$ to account for identical final states. This has to be kept in mind
	when comparing to Ref.~\cite{Bronnum-Hansen:2022tmr} where the gluons were
	not ordered.}
\begin{align}
	\label{eq:rr:xsect}
	\begin{split}
		& \big\langle \FLM{1,2,3,4}{5,6} \big\rangle \equiv
		\\
		& = \int \dq{5} \dq{6} \theta{(E_5 - E_6)} \, \FLM{1,2,3,4}{5,6}.
	\end{split}
\end{align}
Note that we dropped the subscripts indicating the parton type for brevity; we
will continue to use this shortened notation in what follows, unless parton
type becomes relevant. The phase-space element $\dq{k}$ is defined as
\begin{align}
	\dq{k} \equiv \frac{\textrm{d}^{d-1}p_k}{(2\pi)^{d-1}2E_k} \,
	\theta(\emax - E_k) \, ,
\end{align}
where $\emax$ is a parameter that should be equal to or greater than the maximal
energy that a final-state parton can have because of momentum conservation.

To construct the subtraction terms, we need to understand the singularities of
the matrix element in Eq.~\eqref{eq:rr:xsect}. Although, in general, such
singularities can arise when the emitted gluons are either soft or collinear to
other partons, the case of non-factorizable corrections is special \textit{because
	only soft singularities are possible}. However, since we order gluons in energy
and since the matrix element fully factorizes in the double-soft $E_5 \sim E_6
\rightarrow 0$ limit because of the abelian nature of non-factorizable
corrections, it is sufficient to write
\begin{align}
	\label{eq:double-real:regulated}
	\begin{split}
		&\big\langle \FLM{1,2,3,4}{5,6} \big\rangle \\
		&= \big\langle\big[I - S_6\big] \FLM{1,2,3,4}{5,6} \big\rangle \\
		& +\big\langle S_6 \FLM{1,2,3,4}{5,6}\big\rangle \, ,
	\end{split}
\end{align}
to obtain a fully-regulated double-real emission contribution. We remind the
reader that an operator $S_i$ extracts the leading behavior of the function
$F_{\rm LM}^\textrm{nf}$ in the limit where the energy of parton $i$
vanishes, see Ref.~\cite{Caola:2017dug} for additional details.

We now turn our attention to the subtraction term containing the single soft
singularity, i.e. the second term on the right-hand side of
Eq.~\eqref{eq:double-real:regulated}. It is given by
\begingroup
\allowdisplaybreaks
\begin{align}
	&S_6 \FLM{1_q,2_Q,3_{q^\prime},4_{Q^\prime}}{5_g,6_g} = - 2 \,
	g_{s,b}^2 \, \kappa_{qQ} \nonumber \\
	\label{eq:double-real:s6}
	&\eqspace \times \int \dq{6} \theta{(E_5 - E_6)} \,
	\textrm{Eik}_\textrm{nf}(1_q,2_Q,3_{q^\prime},4_{Q^\prime}\,|\, 6_g) \\
	&\eqspace \times \FLM{1_q,2_Q,3_{q^\prime},4_{Q^\prime}}{5_g} \, , \nonumber
\end{align}
\endgroup
where $\kappa_{qQ} = + 1$ if both $q$ and $Q$ are either quarks or anti-quarks,
and $\kappa_{qQ} = -1$ otherwise. The eikonal function in
Eq.~\eqref{eq:double-real:s6} reads
\begin{align}
	\label{eq:rr:eik}
	&\textrm{Eik}_\textrm{nf}(1_q,2_Q,3_{q^\prime},4_{Q^\prime}\,|\, 6_g)
	= \hspace{-5pt} \sum_{\substack{i\in\{1,3\}\\j\in\{2,4\}}}
	\frac{\lambda_{ij}(p_i \cdot p_j)}{(p_i \cdot p_6)(p_j \cdot p_6)} \, ,
\end{align}
with $\lambda_{ij} = +1$ if both $i$ and $j$ are either incoming or outgoing,
and $\lambda_{ij} = -1$ otherwise. We also note that in
Eq.~\eqref{eq:double-real:s6} we have introduced a non-factorizable,
single-gluon emission contribution
\begin{align}
	\label{eq:flm:r}
	\begin{split}
		&\FLM{1,2,3,4}{5} \\
		&\equiv \mathcal{N} \int \textrm{dLips}_{34H} \times \hat{\mathcal{O}}
		{(\{p_{i=1,\dots,5},p_H\})} \\
		&\eqspace \times (2\pi)^d \, \delta^{(d)}{\left(p_1 + p_2 - p_H -
			\sum_{i=3}^5 p_i\right)}\\
		&\eqspace \times 2 \Re{[A_0^1 {A_0^2}^\star]}(1,2,3,4\,|\,5)\,.
	\end{split}
\end{align}

Integration of the eikonal factor over the gluon momentum $p_6$ in
Eq.~(\ref{eq:double-real:s6}) has already been discussed in the literature, see
e.g. Ref.~\cite{Asteriadis:2019dte}. We obtain
\begin{align}
	\label{eq:soft-integrated}
	\begin{split}
		&\big\langle S_6 \FLM{1,2,3,4}{5,6} \big\rangle = - 2 [\alpha_{s,b}] \,
		\kappa_{qQ} \\
		&\eqspace\times \big\langle (2E_5)^{-2\epsilon} \, K_\textrm{nf}(1,2,3,4) \,
		\FLM{1,2,3,4}{5} \big\rangle \, .
	\end{split}
\end{align}
The function $K_\textrm{nf}(1,2,3,4,5)$ can be found in the appendix and
$[\alpha_{s,b}]$ is defined as follows
\begin{align}
	[\alpha_{s,b}] \equiv
	\frac{g_{s,b}^2}{8\pi^2}\frac{(4\pi)^{\epsilon}}{\Gamma(1-\epsilon)} \, .
\end{align}

There is still a soft singularity, $E_5 \rightarrow 0$, in the function
$\FLM{1,2,3,4}{5}$ in Eq.~\eqref{eq:soft-integrated} that needs to be extracted.
Analogously to Eq.~(\ref{eq:double-real:regulated}), we do this by subtracting
and adding the soft limit of gluon $g_5$. We find
\begin{align}
	\label{eq:rr:nlo-regulated}
	\begin{split}
		&\big\langle S_6 \FLM{1,2,3,4}{5,6} \big\rangle = - 2[\alpha_{s,b}] \,
		\kappa_{qQ} \, \big\langle \big[I-S_5\big] \\
		&\eqspace\times (2E_5)^{-2\epsilon} \, K_\textrm{nf}(1,..,4) \,
		\FLM{1,2,3,4}{5} \big\rangle \\
		&\eqspace -2[\alpha_{s,b}] \, \kappa_{qQ}\, \big\langle S_5 \,
		(2E_5)^{-2\epsilon} \, K_\textrm{nf}(1,..,4) \\
		& \eqspace \times \FLM{1,2,3,4}{5} \big\rangle \, .
	\end{split}
\end{align}

The limit of the colour-stripped single-real emission amplitude is similar to
Eq.~\eqref{eq:double-real:s6} and reads
\begin{align}
	\begin{split}
		&S_5 (2 E_5)^{-2\epsilon} \FLM{1,2,3,4}{5} \\
		&= - 2 \, g_{s,b}^2 \, \kappa_{qQ}\;
		(2 E_5)^{-2\epsilon} \textrm{Eik}_\textrm{nf}(1,2,3,4\,|\, 5) \\
		&\eqspace \times \FLMlo{1,2,3,4}\, ,
	\end{split}
\end{align}
where we introduced
\begin{align}
	\begin{split}
		&\FLMlo{1,2,3,4} \\
		&\equiv \mathcal{N} \int \textrm{dLips}_{34H} \times \hat{\mathcal{O}}
		{(\{p_{i=1,\dots,4},p_H\})} \\
		&\eqspace \times (2\pi)^d \, \delta^{(d)}{\left(p_1 + p_2 - p_H - p_3
			- p_4\right)}\\
		&\eqspace \times |A_0|^2(1,2,3,4)\, ,
	\end{split}
\end{align}
to describe the leading-order process.
Upon integration over the unresolved phase space of gluon $g_5$ we find
\begin{align}
	\label{eq:rr:semi-final}
	\begin{split}
		& \big\langle S_5 \, (2E_5)^{-2\epsilon} \, \FLM{1,2,3,4}{5} \big\rangle \\
		&= - [\alpha_{s,b}] (2\emax)^{-4\epsilon}
		\big\langle K_\textrm{nf} \, \FLMlo{1,2,3,4} \big\rangle \, ,
	\end{split}
\end{align}
where we suppressed the dependence of the function $K_\textrm{nf}$ on the Born
momenta.

Finally, we combine Eqs.~(\ref{eq:double-real:regulated},%
~\ref{eq:rr:nlo-regulated},~\ref{eq:rr:semi-final}) and replace
\begin{align}
	\label{eq:rr:renorm}
	[\alpha_{s,b}] \rightarrow \frac{\tilde \alpha_s}{2\pi} \,
	\mu^{2\epsilon} \, ,
\end{align}
where $\tilde \alpha_s = \alpha_s(\mu) e^{\epsilon \gamma_E}/\Gamma(1-\epsilon)$, to
express the result through the strong coupling defined in the
$\overline{\textrm{MS}}$ scheme. The result is the fully-regulated
representation of the double-real contribution to non-factorizable corrections
\begingroup
\allowdisplaybreaks
\begin{align}
	\label{eq:rr:final}
	\begin{split}
		&\big\langle \FLM{1,2,3,4}{5,6} \big\rangle \\
		&= \big\langle\big[I - S_6\big] \FLM{1,2,3,4}{5,6} \big\rangle - 2\,\left (
		\asmu \right ) \, \kappa_{qQ} \\
		& \eqspace \times \big\langle \big[I-S_5\big]\bigg(\frac{2E_5}{\mu}\bigg)^{
			-2\epsilon} K_\textrm{nf} \, \FLM{1,2,3,4}{5} \big\rangle \\
		&+ 2 \, \asmusq \bigg(\hspace{-1.5pt}\frac{2\emax}{\mu}\hspace{-1.5pt}
		\bigg)^{\hspace{-2pt}-4\epsilon} \! \big\langle K_\textrm{nf}^2 \,
		\FLMlo{1,2,3,4} \big\rangle \, . \hspace{-15pt}
	\end{split}
\end{align}
\endgroup

\subsection*{Real-virtual contribution}
\label{sec:comp:real-virtual}

Next, we consider the real-virtual contribution to the NNLO QCD
non-factorizable corrections. It arises from the one-loop corrections to the
process with an additional gluon in the final state
\begin{align}
	\begin{split}
		&q(p_1) + Q(p_2) \\
		&\rightarrow q^\prime(p_3) + Q^\prime(p_4) + g(p_5) + H(p_H) \, .
	\end{split}
\end{align}
The real-virtual contribution to the non-factorizable correction is also
proportional to the colour factor shown in Eq.~(\ref{eq.2}). Hence, following
the discussion of the double-real contribution, we define a
\textit{colour-stripped} amplitude $A_1^i$ as a sum of abelian diagrams where
a virtual gluon is exchanged between the two quark lines and a real gluon is
emitted from line $i$. Using this amplitude, we write the real-virtual
contribution as
\begin{align}
	\label{eq:flv:r}
	\begin{split}
		&\FLV{1,2,3,4}{5} \\
		&\equiv \mathcal{N} \int \textrm{dLips}_{34H} \times \hat{\mathcal{O}}
		{(\{p_{i=1,\dots,5},p_H\})} \\
		&\eqspace \times (2\pi)^d \, \delta^{(d)}{\left(p_1 + p_2 - p_H -
			\sum_{i=3}^5 p_i\right)}\\
		&\eqspace \times 2 \Re{[A_0^1 {A_1^2}^\star + A_0^2 {A_1^1}^\star]}
		(1,2,3,4\,|\,5) \, .
	\end{split}
\end{align}

The only singularity present in $\FLV{1,2,3,4}{5}$ arises in the soft,
$E_5 \to 0$ limit. To regulate it, we write
\begin{align}
	&\big\langle \FLV{1,2,3,4}{5} \big\rangle = \big\langle\big[I - S_5\big]
	\, \FLV{1,2,3,4}{5} \big\rangle \nonumber \\
	\label{eq:real-virtual:regulated}
	& +\big\langle S_5 \, \FLV{1,2,3,4}{5}\big\rangle \, .
\end{align}
Although the first term in the above equation is fully regular inasmuch as the
real emission is concerned, it contains an explicit infrared $1/\epsilon$ pole
which arises as a result of the integration over the loop momentum. We extract
it by writing~\cite{Giele:1991vf,*Kunszt:1994np,*Catani:1996jh,*Catani:1996vz}
\begin{align}
	\label{eq:rv:virtpole5pt}
	\begin{split}
		&\FLV{1,2,3,4}{5} = \asmu \, 2 \, \kappa_{qQ} \, I_1(\epsilon) \,
		\FLM{1,2,3,4}{5} \\
		&+ \FLVfin{1,2,3,4}{5} \, ,
	\end{split}
\end{align}
where
\begin{align}
	\label{eq:rv:catani-like}
	I_1(\epsilon) \equiv \frac{1}{\epsilon} \ln{\left(\frac{p_1 \cdot p_4 \ p_2
			\cdot p_3}{p_1 \cdot p_2 \ p_3 \cdot p_4}\right)} \, ,
\end{align}
$\FLM{1,2,3,4}{5}$ is the colour-stripped single-real emission contribution
defined in Eq.~\eqref{eq:flm:r} and $\FLVfin{1,2,3,4}{5}$ is the
$\mathcal{O}(\epsilon^0)$ coefficient in the $\epsilon$-expansion of
Eq.~\eqref{eq:flv:r}.

We now discuss the second term on the right-hand side of
Eq.~\eqref{eq:real-virtual:regulated}. The soft-gluon limit of any one-loop QCD
amplitude is known~\cite{Bern:1999ry,*Kosower:1999rx,*Catani:2000pi}. It
contains two terms -- the product of the tree-level eikonal current and a
one-loop amplitude without the soft gluon, as well as the product of a one-loop
correction to the eikonal current and the relevant tree-level amplitude. Since
the one-loop correction to the eikonal current is purely non-abelian, it plays
no role in the computation of non-factorizable corrections. We discard it and
write
\begin{align}
	\begin{split}
		&S_5 \FLV{1,2,3,4}{5} \\
		&= - 2 \, g_{s,b}^2 \, \kappa_{qQ} \int \dq{5} \, \textrm{Eik}_\textrm{nf}
		(1,2,3,4\,|\, 5) \\
		&\eqspace\times \FLVlo{1,2,3,4},
		\label{eq26}
	\end{split}
\end{align}
where we introduced
a colour-stripped one-loop virtual contribution
\begin{align}
	\begin{split}
		&\FLVlo{1,2,3,4} \\
		&\equiv \mathcal{N} \int \textrm{dLips}_{34H} \times \hat{\mathcal{O}}
		{(\{p_{i=1,\dots,4},p_H\})} \\
		&\eqspace \times (2\pi)^d \, \delta^{(d)}{\left(p_1 + p_2 - p_H - p_3 -
			p_4\right)}\\
		&\eqspace \times 2\,\Re{[A_0 A_1^\star]} (1,2,3,4) \, .
	\end{split}
\end{align}

The integral over unresolved momentum $p_5$ in Eq.~(\ref{eq26}) evaluates to
\begin{align}
	\label{eq:rv:intsub}
	\begin{split}
		&\big\langle S_5 \FLV{1,2,3,4}{5}\big\rangle = - 2 \, \kappa_{qQ} \, \asmu
		\left(\frac{{2\emax}}{\mu}\right)^{-2\epsilon} \\
		&\eqspace\times \big\langle K_\textrm{nf}(1,2,3,4) \, \FLVlo{1,2,3,4}
		\big\rangle \, .
	\end{split}
\end{align}

To proceed further, we note that $\FLVlo{1,2,3,4} $ contains infrared poles
from the loop integration. We make them explicit by writing
\begin{align}
	\label{eq:rv:virtpole4pt}
	\begin{split}
		&\FLVlo{1,2,3,4} =\asmu \,2 \, \kappa_{qQ} \, I_1(\epsilon) \,
		\FLMlo{1,2,3,4} \\
		& + \FLVfinlo{1,2,3,4} \, .
	\end{split}
\end{align}
The function $I_1(\epsilon)$ has already appeared in
Eq.~\eqref{eq:rv:catani-like}.

Combining Eqs.~(\ref{eq:real-virtual:regulated},~\ref{eq:rv:virtpole5pt},%
~\ref{eq:rv:intsub},~\ref{eq:rv:virtpole4pt}), we obtain the final result
for the real-virtual contribution to the non-factorizable corrections
\begingroup
\allowdisplaybreaks
\begin{align}
	\label{eq:rv:final}
	\begin{split}
		&\big\langle \FLV{1,2,3,4}{5} \big\rangle \\
		&= \asmu \, \kappa_{qQ} \big\langle 2 \, I_1(\epsilon) \, \big[I - S_5\big]
		\FLM{1,2,3,4}{5} \big\rangle \\
		& + \big\langle \big[I - S_5\big]\FLVfin{1,2,3,4}{5} \big\rangle \\
		& - 4 \, \asmusq \left(\frac{2\emax}{\mu}\right)^{-2\epsilon} \! \big\langle
		I_1(\epsilon) K_\textrm{nf}\, \FLMlo{1,2,3,4} \big\rangle \\
		& - 2 \, \asmu \, \kappa_{qQ} \left(\frac{2\emax}{\mu}\right)^{-2\epsilon}
		\! \big\langle K_\textrm{nf} \, \FLVfinlo{1,2,3,4} \big\rangle \, .
	\end{split}
\end{align}
\endgroup

\subsection*{Double-virtual contribution}
\label{sec:comp:double-virtual}

The last contribution that we need to consider is the two-loop non-factorizable
correction to the process
\begin{align}
	q(p_1)+Q(p_2)\rightarrow q^\prime(p_3) + Q^\prime(p_4) + H(p_H) \, .
\end{align}
We write the two-loop amplitude of this process separating the $1/\epsilon$ infrared
poles from the finite remainder using the results in Refs.~\cite{Catani:1998bh}.
Since the non-factorizable corrections are abelian, the divergent structure of
the two-loop amplitude is fully determined by the square of $I_1(\epsilon)$, c.f.
Eq.~\eqref{eq:rv:catani-like}. We write
\begin{align}
	\label{eq:vv:final}
	\begin{split}
		& \big\langle \FLVV{1,2,3,4} \big\rangle
		= \asmusq \big\langle 2 \, I_1(\epsilon)^2 \FLMlo{1,2,3,4} \big\rangle\\
		&\eqspace + \asmu \, \kappa_{qQ} \, \big\langle 2 \,  I_1(\epsilon) \,
		F_\textrm{LV,fin}^{\textrm{nf}}(1,2,3,4) \big\rangle \\
		&\eqspace + \big\langle F_\textrm{LVV,fin}^{\textrm{nf}}(1,2,3,4)\big\rangle \, ,
	\end{split}
\end{align}
where $F_\textrm{LVV,fin}^{\textrm{nf}}$ is the finite result for the two-loop
amplitude.

\subsection*{Explicit pole cancellation and IR finite result}
\label{sec:comp:pole-cancellation}

The final result for the cross section is obtained by combining the
double-real, real-virtual and double-virtual contributions given in
Eq.~\eqref{eq:rr:final}, Eq.~\eqref{eq:rv:final} and Eq.~\eqref{eq:vv:final},
respectively. We write the partonic cross section as
\begingroup
\allowdisplaybreaks
\begin{align}
	& {\rm d}\sigma^\textrm{nf}_\textrm{nnlo} = \frac{ T_R^2 (N_c^2-1) }{2s}\,
	\Big[ \big \langle \FLM{1,2,3,4}{5,6} \big\rangle \nonumber \\
	&\eqspace + \big\langle \FLV{1,2,3,4}{5} \big\rangle + \big\langle
	\FLVV{1,2,3,4} \big\rangle \Big] \nonumber \\
	&= \frac{ T_R^2 (N_c^2-1) }{2s} \, \bigg[\big\langle\big[I - S_6\big]
	\FLM{1,2,3,4}{5,6} \big\rangle \nonumber \\
	\begin{split}
		\label{eq:final:res}
		&- 2\, \asmu \big\langle \big[I-S_5\big] \, \mathcal{W}(E_5;1,..,4) \,
		\FLM{1,2,3,4}{5} \big\rangle \\
		&+ 2 \, \asmusq\big\langle \mathcal{W}(\emax;1,..,4)^2 \FLMlo{1,2,3,4}
		\big\rangle
	\end{split} \\
	&+ \big\langle\big[I-S_5\big]\FLVfin{1,2,3,4}{5} \big\rangle \nonumber \\
	&- 2 \, \asmu \big\langle \mathcal{W}(\emax;1,..,4) \, \FLVfinlo{1,2,3,4}
	\big\rangle \nonumber \\
	&+ \big\langle F_\textrm{LVV,fin}^{\textrm{nf}}(1,2,3,4)\big\rangle \,
	\bigg ] \nonumber \, .
\end{align}
\endgroup
In Eq.~(\ref{eq:final:res}) we introduced a finite function ${\cal
	W}(E;1,2,3,4)$ defined as\footnote{The $\epsilon$-expansion of function
	$K_\textrm{nf}$ can be found in the appendix, see
	Eq.~\eqref{eq:app:intsub:expansion}.}
\begin{align}
	\label{eq:Wdef}
	\begin{split}
		&\mathcal{W}(E;1,2,3,4) \equiv \kappa_{qQ} \bigg[ \left(\frac{2 E}{\mu}
		\right)^{-2\epsilon} \! K_\textrm{nf}(\epsilon) - \textrm{I}_1(\epsilon)
		\bigg] \\
		& = \kappa_{qQ} \bigg[ -2 \ln{\left(\frac{2E}{\mu}\right)} \ln{\left(
			\frac{p_1\cdot p_4\ p_3 \cdot p_2}{p_1 \cdot p_2 \ p_3 \cdot p_4}\right)}\\
		&+ \hspace{-3pt} \sum_{\substack{i\in\{1,3\}\\j\in\{2,4\}}} \hspace{-3pt}
		\lambda_{ij} \left(\frac{1}{2} \ln^2(\eta_{ij}) + \textrm{Li}_2
		(1-\eta_{ij}) \right) \bigg] + \mathcal{O}(\epsilon) \, ,
	\end{split}
\end{align}
where $\eta_{ij} = 1 - \cos \theta_{ij}$ with angles defined in the partonic
centre-of-mass frame. The representation of the partonic cross section given in
Eq.~(\ref{eq:final:res}) makes the cancellation of all $1/\epsilon$ poles
manifest and allows us to take the $\epsilon \rightarrow 0$ limit right away.
Note that upon doing so, the coupling constant $\tilde \alpha_s$ becomes
$\alpha_s(\mu)$, the standard $\overline {\rm MS}$ coupling constant.

\section{Numerical implementation}
\label{sec:implementation}

The numerical implementation of the non-factorizable contribution
Eq.~\eqref{eq:final:res} requires double-real amplitudes as well as finite parts
of real-virtual amplitudes and double-virtual amplitudes. To obtain the required
double-real amplitudes, we extend the calculation of the factorizable NNLO QCD
corrections reported in Ref.~\cite{Asteriadis:2021gpd}.

To compute the real-virtual contributions, we require non-factorizable one-loop
amplitudes for the processes $q+Q\rightarrow q^\prime + Q^\prime + H$ and
$q+Q\rightarrow q^\prime + Q^\prime + H+g$. These amplitudes were computed in
Ref.~\cite{Campanario:2013fsa} and we employ them in our numerical
implementation. Extracting the non-factorizable contribution
from the existing code requires only minor changes.\footnote{We are grateful to T.~Figy for making the code used for the computations reported in Ref.~\cite{Campanario:2013fsa} available to us.} However, it turns out to be
non-trivial to achieve stable and reliable numerical results close to
singular limits.

The existing implementation uses on-the-fly numerical Passarino-Veltman
reduction and the \texttt{OneLOop} library~\cite{vanHameren:2010cp} for the
evaluation of scalar integrals. To reach sufficient numerical accuracy we limit
catastrophic cancellation by working with scaleless $\mathcal{O}(1)$ quantities.
This is achieved by scaling out the energy of the incoming partons in all
momenta and masses in each phase space point and re-introducing it at the very
end of the calculation.

Furthermore, we find it necessary to work with quadruple precision. With these
two measures we achieve agreement with the infrared pole prediction in
Eq.~\eqref{eq:rv:catani-like} to more than 10 digits for most phase space
points. In addition to checking the amplitude's pole structure, we also find a
satisfactory agreement between the exact six-point amplitude and its expected
limit when the energy of the final-state gluon becomes small, see
Eq.~\eqref{eq26}. Obviously, this last feature is a necessary requirement for
being able to use Eq.~(\ref{eq:final:res}) for phenomenological studies.

For the finite remainder of the two-loop amplitude,
$F_\textrm{LVV,fin}^{\textrm{nf}}$, we use the results of
Ref.~\cite{Liu:2019tuy}.
These results are obtained in the eikonal approximation
which provides the leading term in the expansion of this amplitude in $p_\perp /
\sqrt{s}$ where $p_\perp$ is a typical transverse momentum of the final-state
tagging jets. This approximation is motivated by typical WBF signatures and the
fiducial selection cuts derived from them.\footnote{We note that fully analytic result for the leading eikonal approximation are available in Ref.~\cite{Gates:2023iiv}.}

As a final comment we note that the finite part of the two-loop amplitude
~\cite{Liu:2019tuy} that we use in this computation is an approximation to the
exact result which, so far, remains unknown. In particular, the two-loop
amplitude computed in the eikonal approximation \cite{Liu:2019tuy} is infrared
finite which means that there is no connection between the first two terms on
the right-hand side of Eq.~\eqref{eq:vv:final}, required to cancel divergences
in the double-real and real-virtual contributions, and $\big\langle
F_\textrm{LVV,fin}^{\textrm{nf}}(1,2,3,4)\big\rangle$. However, as we will show
in Section~\ref{sec:results}, it is quite unlikely that the missing parts of the
finite remainder of the two-loop amplitude that are linked to the cancellation
of infrared divergences can impact the phenomenology of weak boson fusion in a
significant way.

\section{Results}
\label{sec:results}

\begin{figure*}
	\centering
	\includegraphics[height=270pt]{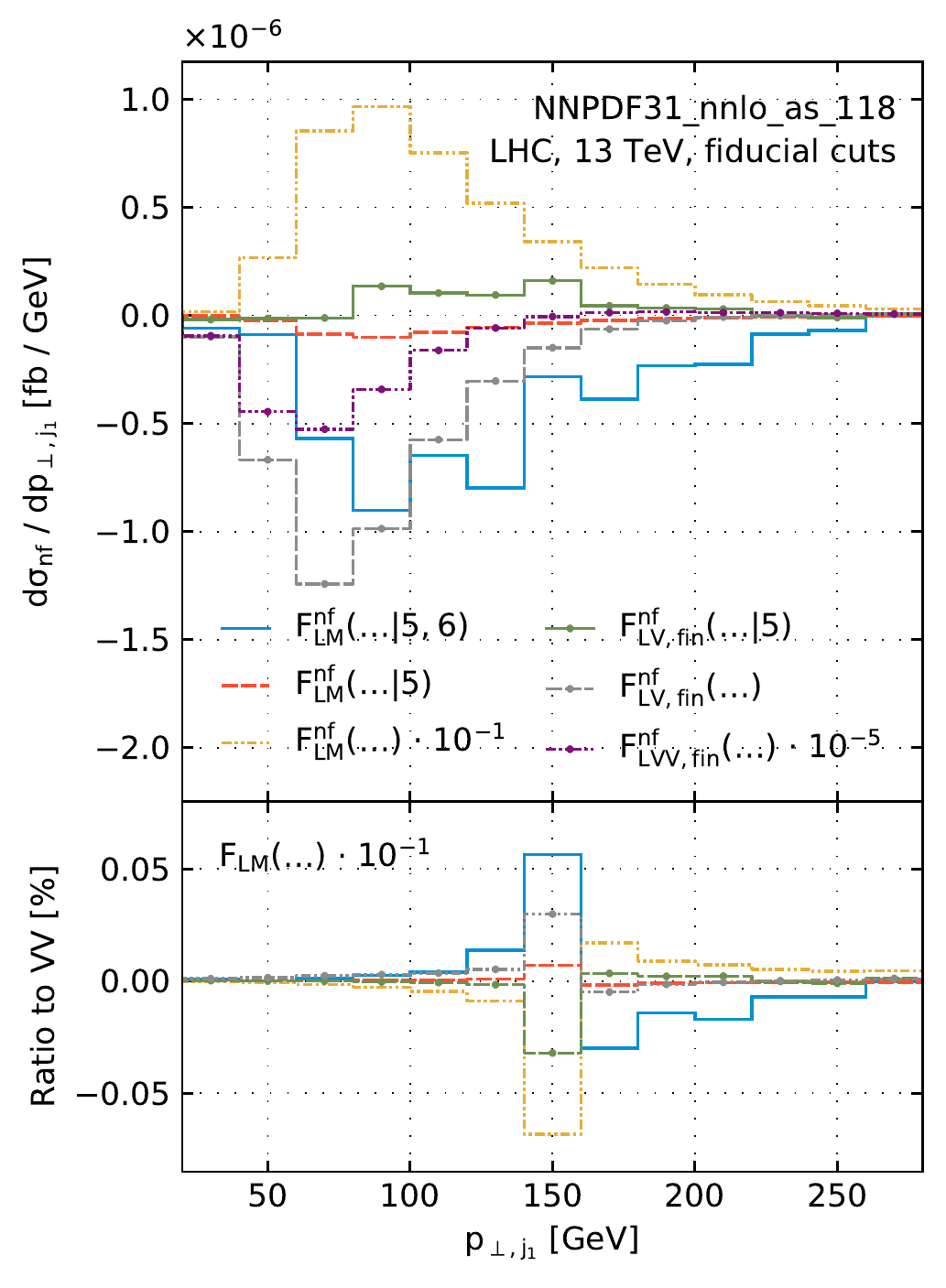} \hspace{40pt}
	\includegraphics[height=270pt]{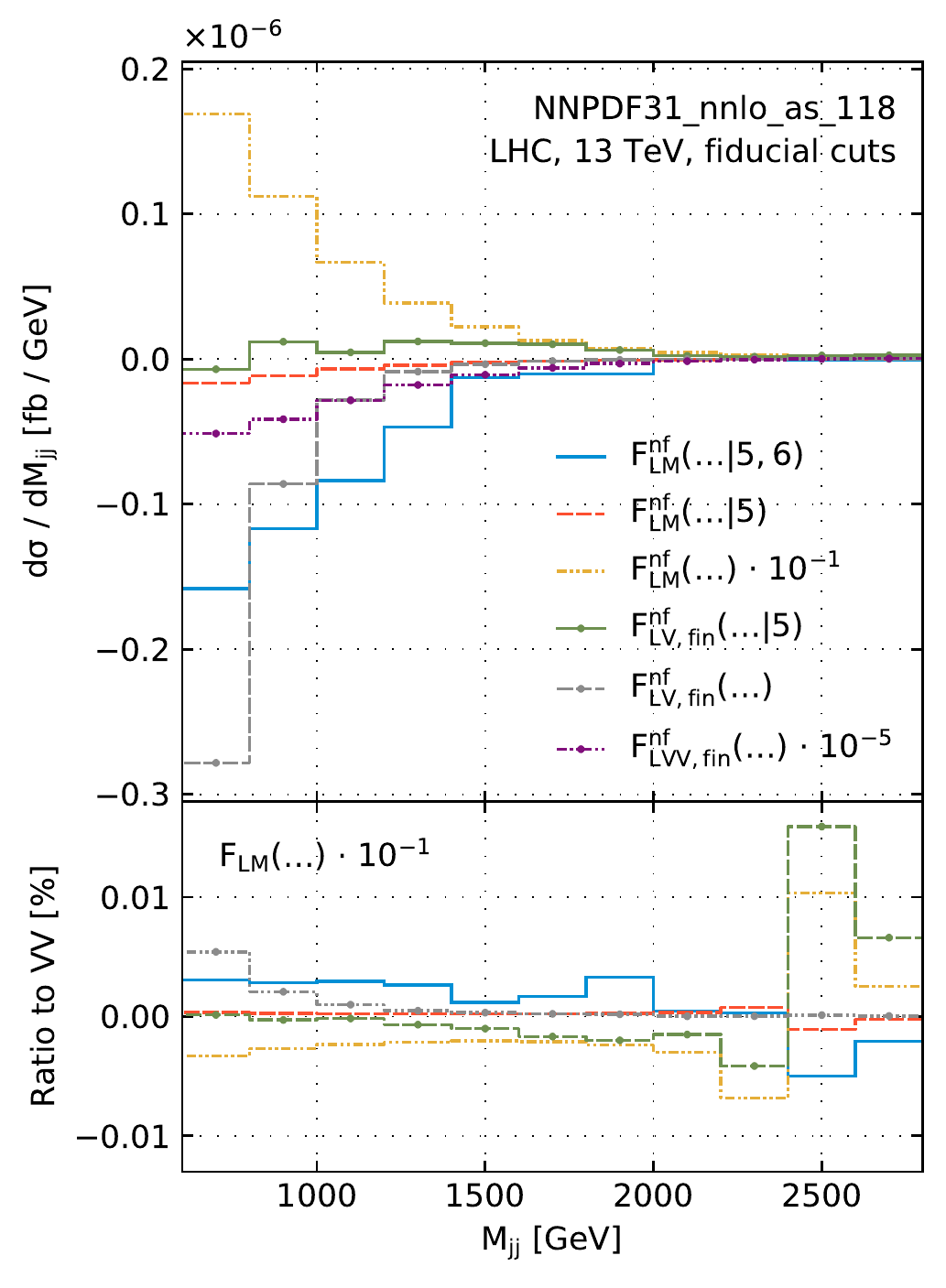} \hspace{25pt}
	\caption{Non-factorizable contribution to the transverse momentum
		distributions of the leading jet (left)
		and to the distribution of the invariant mass of the tag-jet system
		(right). Contributions are shown individually for different terms on the
		right-hand side of Eq.~\eqref{eq:final:res} and we label them with the
		present matrix element, e.g. the plot label $F_\textrm{LM}^\textrm{nf}(1,
		2,3,4 \, | \, 5)$ refers to the contribution of the \textit{full} second
		term. Note that, in the plots we use ellipses for the sequence of Born
		momenta, $1,2,3,4$, for representational purposes. For each plot (and
		differently in upper and lower panes) contributions are scaled to be of
		similar orders. The lower pane shows the ratio with respect to
		double-virtual contributions. See text for further details.}
	\label{histo:ptj}
\end{figure*}

The goal of this section is to compute the non-factorizable NNLO QCD corrections
to Higgs boson production in weak boson fusion and to compare them to the factorizable
ones. To do that, we adopt standard parameters and kinematic selection
criteria from Refs.~\cite{Asteriadis:2019dte,Asteriadis:2022ebf}; we reproduce
them here for completeness.

We consider $13$ TeV proton-proton collisions. The Higgs boson is chosen to be
stable with a mass of $m_H =125~{\rm GeV}$. Vector boson masses are taken to be
$M_W= 80.398~{\rm GeV}$ and $M_Z = 91.1876~{\rm GeV}$ with widths $\Gamma_W =
2.105~{\rm GeV}$ and $\Gamma_Z = 2.4952~{\rm GeV}$, respectively. Weak couplings
are derived from the Fermi constant $G_F = 1.16639 \times 10^{-5} {\rm GeV}^{
	-2}$ and the CKM matrix is set to the identity matrix.

We use \texttt{NNPDF31-nnlo-as-118} parton distribution
functions~\cite{NNPDF:2017mvq} and $\alpha_s(M_Z) = 0.118$ for all
calculations reported below. The evolution of both parton distribution functions
and the strong coupling constant is obtained directly from
LHAPDF~\cite{Buckley:2014ana}. The dynamical renormalization and factorization
scales are set equal, $\mu_R = \mu_F = \mu$, with the central
value~\cite{Cacciari:2015jma}
\begin{align}
	\mu_0 = \sqrt{ \frac{m_H}{2} \sqrt{
			\frac{m_H^2}{4} + p_{\perp,H}^2} } \, .
	\label{eq:dynsc}
\end{align}

To define the WBF fiducial volume we employ the inclusive anti-$k_\perp$ jet
algorithm \cite{Cacciari:2008gp} with $R = 0.4$. Events are required to contain
at least two jets with transverse momenta $p_{\perp,j} > 25~{\rm GeV}$ and
rapidities $|y_j| < 4.5$. The two leading-$p_\perp$ jets must have
well-separated rapidities, $|y_{j_1} - y_{j_2}| > 4.5$, and their invariant mass
should be larger than $600~{\rm GeV}$. In addition, the two leading jets must be
in separate hemispheres in the laboratory frame; this is enforced by requiring
that the product of their rapidities in the laboratory frame is negative,
$y_{j_1} y_{j_2} < 0$.

The analysis of the double-virtual contribution to the non-factorizable
correction to Higgs boson production in weak boson fusion has already been
performed in Refs.~\cite{Liu:2019tuy,Dreyer:2020urf}. The new elements that we
add to this analysis are the double-real and real-virtual
contributions. Although typically one expects that all types of contributions
are comparable in magnitude, we find that for Higgs production in WBF this is
not the case.

For example, computing the non-factorizable NNLO QCD corrections to the fiducial
WBF cross section for central values of the renormalization and factorization
scales and for values of parameters as described above, we find
\begin{equation}
	\sigma_\textrm{nf} = -3.1~{\rm fb}\, .
	\label{eq36}
\end{equation}
We note that this result has a significant scale uncertainty because
non-factorizable corrections appear at NNLO \emph{for the very first time} and
there is no mechanism to e.g. compensate the change in the strong coupling
constant when the renormalization scale is modified. For this reason it is not
surprising that we find ${\cal O}(40\, \%)$ uncertainty in $\sigma_{\rm nf}$
upon varying $\mu_R$ and $\mu_F$ within an interval $[\mu_0/2,~2\,\mu_0]$. We also
note that $\sigma_\textrm{nf}$ provides ${\cal O}(0.5)$ percent correction to
the fiducial cross section computed through NNLO QCD in the factorization
approximation~\cite{Asteriadis:2021gpd} and is about a factor of ten smaller
than the factorizable NNLO QCD corrections.

As we already mentioned, one would normally expect that double-virtual,
real-virtual and real-real corrections provide comparable contributions to
$\sigma_\textrm{nf}$. However, it turns out that this is not the case and that
\textit{only} $0.01$ percent of $\sigma_\textrm{nf}$ comes from the real-virtual
and the double-real contributions whereas the dominant $99.99$ percent comes
from the double-virtual one.

This relation between the double-virtual and all the other contributions holds
for all kinematic distributions that we considered. To give some examples, in
Fig.~\ref{histo:ptj} we show the different contributions to the transverse
momentum distributions of the hardest jet and the
distribution of the invariant mass of the pair of leading jets.

To understand the reason for this unusual suppression of the double-real and the
real-virtual contributions, consider the quantity
\begin{align}
	\label{eq:L}
	{L}(1,2,3,4) = \ln{\left(\frac{p_1 \cdot p_4 \ p_3 \cdot p_2}{p_1 \cdot p_2
			\ p_3 \cdot p_4}\right)} \, ,
\end{align}
which arises upon integration of the eikonal current describing single gluon
emission. We note that this quantity appears in the integrated subtraction
term described by the function $\mathcal{W}(E;1,2,3,4)$ defined in
Eq.~\eqref{eq:Wdef}.

For instance, to estimate the contribution of two soft gluons to the non-factorizable
corrections in the presence of fiducial WBF cuts, we consdier the following
integral
\begin{align}
	\sigma_{RR} \sim \asmusq N_c^2 \big\langle L^2(1,2,3,4) \,
	F^{\rm nf}_{\rm LM}(1_q,2_q,3_q,4_q) \big\rangle \, . \hspace{-1pt}
\end{align}

To proceed we use the fact that in the relevant phase-space region $p_3$ and
$p_4$ are nearly collinear to $p_1$ and $p_2$, respectively, and compute the
function $L$ in this limit. To this end, we write
\begin{align}
	\begin{split}
		& p_3 = \alpha_3 \, p_1 + \beta_3 \, p_2 + p_{3,\perp} \, , \\
		& p_4 = \alpha_4 \, p_1 + \beta_4 \, p_2 + p_{4,\perp} \, ,
	\end{split}
\end{align}
where $\alpha_3, \beta_4 \sim 1$ and
\begin{align}
	p_{i,\perp} \cdot p_1 = p_{i,\perp} \cdot p_2 = 0 \, ,
\end{align}
for $i \in \{3,4\}$. From the mass-shell condition for outgoing quarks, we
obtain
\begin{align}
	\beta_3 \sim \frac{p_{3,\perp}^2}{s} \ll 1 \, , \quad \alpha_4 \sim
	\frac{p_{4,\perp}^2}{s} \ll 1 \, .
\end{align}
We thus find
\begin{align}
	\begin{split}
		\label{eq:l:approx}
		L(1,2,3,4)
		& = -\ln \left( 1 + \frac{\beta_3 \, \alpha_4}{\alpha_3 \, \beta_4} -
		\frac{2 \, \vec{p}_{3,\perp} \!\cdot \vec{p}_{4,\perp}}{s \, \alpha_3 \,
			\beta_4} \right) \\
		& \approx \frac{2\vec p_{3,\perp} \! \cdot \vec p_{4,\perp}}{s} \, .
	\end{split}
\end{align}
A typical transverse momentum in Higgs production in weak boson fusion is $\sim
60~{\rm GeV}$ and a typical partonic centre-of-mass energy is approximately
$\sqrt{s} \approx 600~{\rm GeV}$. Therefore, $L \sim 10^{-2}$ in the relevant
region of the partonic phase space and we find
\begin{align}
	\begin{split}
		\sigma_{RR} &\sim \asmusq N_c^2 \big\langle L^2(1,2,3,4)
		F^{\rm nf}_{\rm LM}(1_q,2_q,3_q,4_q) \big\rangle\\
		& \sim \asmusq \; 10^{-4}\; \sigma_{\rm LO} \, ,
	\end{split}
\end{align}
where we used $N_c^2 \big\langle F^{\rm nf}_{\rm LM}(1_q,2_q,3_q,4_q)
\big\rangle = \sigma_{\rm LO}$.

In comparison, virtual corrections do not vanish in the forward region. In
fact, as shown in Ref.~\cite{Liu:2019tuy}, they are characterised by a
phase-space dependent function $ \chi_{\rm nf}$ which is ${\cal O}(\pi^2)$ in
the forward region. We then estimate
\begin{align}
	\begin{split}
		\sigma_{VV} & \sim
		\asmusq N_c^2
		\big\langle \; \chi_{\rm nf}(1,2,3,4) \; F^{\rm nf}_{\rm LM}(1,2,3,4)
		\big\rangle \\
		& \approx \asmusq \, 10 \, \sigma_{\rm LO} \, ,
	\end{split}
\end{align}
where we used $\pi^2 \approx 10$.
Taking the ratio, we obtain
\begin{align}
	\frac{\sigma_{RR}}{\sigma_{VV}} \sim 10^{-5} \, ,
\end{align}
which is consistent with the results of the explicit computation presented
earlier in this section.

We have checked that the extraordinarily strong suppression of the double-real
and real-virtual corrections is a consequence of the fiducial cuts which are
used to identify events when the Higgs boson is produced in weak boson fusion.
If the cuts are relaxed so that one does not require strong rapidity separation
of the two tagging jets and a strong constraint on their invariant mass, the
double-real and real-virtual contributions increase by several orders of
magnitude. In fact, they become comparable to the double-virtual corrections
which only grows by an ${\cal O}(1)$ factor.

\section{Conclusions}
\label{sec:conclusions}

In this paper we extended the calculation of non-factorizable contributions to
Higgs boson production in weak boson fusion at $\mathcal{O}(\alpha_s^2)$ by
combining the results for the double-virtual contributions in the eikonal
approximation~\cite{Liu:2019tuy} with non-factorizable real-virtual and
double-real QCD corrections. We observed that, thanks to the fiducial cuts used
to identify WBF events, and a peculiar enhancement of the double-virtual
contributions, the non-factorizable NNLO QCD corrections are entirely dominated
by two-loop virtual effects. We have checked that the striking dominance of the
two-loop virtual corrections extends to all major kinematic distributions
relevant for Higgs production in WBF.

Outside the fiducial region the relative importance of the various contributions
levels out. However, the eikonal approximation will also start to break down. It
would, therefore, be interesting to understand how to go beyond the eikonal approximation
for the double-virtual amplitude and estimate the impact of non-vanishing
transverse momenta of the final-state jets on the two-loop correction. This
question may be of some relevance for studies that select harder Higgs bosons
which happens, for example, when one considers Higgs decays into a $b$-quark
pair. We leave this question for future investigations.

\section*{Acknowledgments}

We thank S.~Pl\"atzer for useful conversations.
We are grateful to T.~Figy for providing a Fortran code to compute the
one-loop amplitudes for the $q \ Q \to q' \ Q' + H + g$ process and for explaining to us how to use it. This research is
partially supported by the Deutsche Forschungsgemeinschaft (DFG, German Research
Foundation) under grant 396021762~-~TRR~257. The research of K.A. is supported
by the United States Department of Energy under Grant Contract DE-SC0012704. The
research of C.B.H. is supported by the Carlsberg Foundation.

\onecolumngrid
\appendix*

\section{Integrated soft eikonal}

In this appendix we present results for the integrated soft eikonal function
that we have written in terms of the function $K_\textrm{nf}$, c.f.
Eq.~\eqref{eq:soft-integrated}. The exact form of $K_\textrm{nf}$ reads
\begin{align}
	\label{eqa1}
	\begin{split}
		&K_\textrm{nf}(1_q,2_Q,3_{q^\prime},4_{Q^\prime};\epsilon) = \frac{1}
		{\epsilon^2}\bigg[\frac{\Gamma^2(1-\epsilon)}{\Gamma(1-2\epsilon)} \bigg]
		\sum_{\substack{i\in\{1,3\}\\j\in\{2,4\}}} \lambda_{ij} \, \eta_{ij}
		\, {}_2\textrm{F}_1(1,1;1-\epsilon;1-\eta_{ij}) \, ,
	\end{split} \nonumber \\[-25pt]
\end{align}
where we use $\eta_{ij} \equiv 1 - \cos\theta_{ij} \equiv (p_i \cdot p_j) /
(2E_iE_j)$.

It may appear from Eq.~\eqref{eqa1} that the function $K_\textrm{nf}$
contains second-order poles in $\epsilon$. This, however, cannot be the case
since collinear singularities cannot appear in non-factorizable contributions.
An explicit computation yields the result that confirms this expectation.
Expanding $K_{\rm nf}$ in $\epsilon$, we obtain
\begin{align}
	\label{eq:app:intsub:expansion}
	\begin{split}
		K_\textrm{nf}(1_q,2_Q,3_{q^\prime},4_{Q^\prime};\epsilon) &= \frac{1}
		{\epsilon} \ln{\left(\frac{p_1 \cdot p_4 \ p_3 \cdot p_2}{p_1 \cdot p_2 \
				p_3 \cdot p_4}\right)} + \sum_{\substack{i\in\{1,3\}\\j\in\{2,4\}}}
		\lambda_{ij} \left(\frac{1}{2} \ln^2(\eta_{ij}) + \textrm{Li}_2
		(1-\eta_{ij}) \right) + \mathcal{O}(\epsilon) \, .
	\end{split}
\end{align}

\twocolumngrid
\let\oldaddcontentsline\addcontentsline
\renewcommand{\addcontentsline}[3]{}
\bibliography{nf}{}
\let\addcontentsline\oldaddcontentsline
	
\end{document}